\documentclass[iop]{emulateapj}
\usepackage{graphicx}
\usepackage{color}

\shorttitle{H to Zn Ionization Equilibrium for the $\kappa$-distributions}
\shortauthors{Dzif\v{c}\'akov\'a et al.}

\begin{document}

\title{H to Zn Ionization Equilibrium for the Non-Maxwellian Electron $\kappa$-distributions: Updated Calculations}

\author{E. Dzif\v{c}\'akov\'a\altaffilmark{1}} 
\affil{Astronomical Institute of the Academy of Sciences of the Czech Republic, Fri\v{c}ova 298, 251 65 Ond\v{r}ejov, Czech Republic}
\author{J. Dud\'ik\altaffilmark{1}}
\affil{Newton International Fellow, DAMTP, CMS, University of Cambridge, Wilberforce Road, Cambridge CB3 0WA, United Kingdom}
\email{elena@asu.cas.cz}

\altaffiltext{1}{DAPEM, Faculty of Mathematics Physics and Computer Science, Comenius University, Mlynsk\'a Dolina F2, 842 48 Bratislava, Slovakia}

\begin{abstract}
New data for calculation of the ionization and recombination rates have have been published in the past few years. Most of these are included in CHIANTI database. 
We used these data to calculate collisional ionization and recombination rates for the non-Maxwellian $\kappa$-distributions with an enhanced number of particles in the high-energy tail, which have been detected in the solar transition region and the solar wind. Ionization equilibria for elements H to Zn are derived. The $\kappa$-distributions significantly influence both the ionization and recombination rates and widen the ion abundance peaks. In comparison with Maxwellian distribution, the ion abundance peaks can also be shifted to lower or higher temperatures. The updated ionization equilibrium calculations result in large changes for several ions, notably Fe VIII--XIV. The results are supplied in electronic form compatible with the CHIANTI database. 
\end{abstract}


%
%
\section{Introduction}
\label{Sect:1}


One of the most widely used assumptions in the interpretation of astrophysical spectra is that the emitting system is in thermal equilibrium. This means that the distribution of particle energies is at least locally Maxwellian, and can be characterized by the Boltzmann-Gibbs statistics which has one parameter, the temperature.

The generalization of the Bolzmann-Gibbs statistics proposed by \citet{Tsallis88,Tsallis09}, results in $\kappa$-distributions \citep[e.g.,][]{Leubner02,Leubner04a,Leubner04b,Collier04,Livadiotis09}, characterized by a parameter $\kappa$ and exhibiting a near-Maxwellian core and a high-energy power-law tail (Sect. \ref{Sect:2}). 
First proposed by \citet{Vasyliunas68}, the $\kappa$ distributions are now used to fit the observations of a wide variety of astrophysical environments, e.g. in-situ measurements of particle distributions in planetary magnetic environments \citep[e.g.,][]{Pierrard96,Mauk04,Schippers08,Xiao08,Dialynas09} and solar wind \citep[e.g.,][]{Collier96,Maksimovic97a,Maksimovic97b,Pierrard99,Nieves08,LeChat09,LeChat11,Pierrard12}, as well as photon spectra of solar flare plasmas \citep[e.g.,][]{Kasparova09,Oka13}, emission line spectra of planetary nebulae and galactic sources (\citeauthor{Nicholls12}\,\citeyear{Nicholls12}; see also \citeauthor{Binette12}\,\citeyear{Binette12}) and even solar transition region \citep{Dzifcakova11}. A review on the $\kappa$-distributions and their applications in astrophysical plasma can be found, e.g., in \citet{Pierrard10}.

In the solar corona, presence of the $\kappa$-distributions, or distributions exhibiting high-energy tails, can be expected due to particle acceleration processes arising as a result of ``nanoflare'' heating. The nanoflares are an unknown energy release process of impulsive nature, occuring possibly in storms heating the solar corona \citep[e.g.,][]{Tripathi10,Viall12,Bradshaw12,Winebarger12}. While the direct evidence for enhanced suprathermal populations in the solar corona is still lacking \citep{Feldman07,Hannah10}, \citet{Pinfield99} reported that the intensities of the \ion{Si}{3} transition region lines observed by the SOHO/SUMER instrument do not correspond to a single Maxwellian distribution. Using their data, \citet{Dzifcakova11Si} showed that the observed intensities can be explained by $\kappa$-distributions once the photoexcitation is taken into account. These authors diagnosed $\kappa$\,=\,7 in the active region observed on the solar limb. Higher values of $\kappa$ were diagnosed for the quiet 
Sun and coronal hole, indiciating that the departures from the Maxwellian distribution can be connected to the local magnetic activity. The diagnostic method also works for inhomogeneous plasmas characterized by differential emission measure. That the $\kappa$-distributions can be present in the solar corona is also suggested by their presence in the solar wind \citep[e.g.,][]{Pierrard99,Vocks03}.

Direct diagnostics of $\kappa$-distributions in the solar corona using extreme-ultraviolet lines observed by the Hinode/EIS spectrometer \citep{Culhane07} were attempted by \citet{Dzifcakova10} and \citet{Mackovjak13}. These authors proposed methods for simultaneous diagnostics of the plasma temperature, electron density, and $\kappa$. However, majority of the line ratios sensitive to $\kappa$-distributions suffer from poor photon statistics, errors in atomic data and/or plasma inhomogeneities. In spite of this, one of the main results of these works is that the sensitivity to $\kappa$-distributions, or to departures from the Maxwellian distribution in general, is enhanced if the line ratios involve lines originating in neighbooring ionization stages.
This is due the sensitivity of the line emissivity to the abundance of the emitting ion that depends directly on the ionization equilibrium in turn highly dependent on the type of the distribution \citep[e.g.,][]{Dzifcakova02,Wannawichian03, Dzifcakova06}.
%
%

In the past decade, new calculations of the ionization, recombination rates and ionization equilibirium for the Maxwellian electron distribution were published. These are summarized in the continually updated CHIANTI database, currently available in version 7.1 \citep[e.g.,][]{Dere97,Dere09,Landi12,Landi13}. These new calculations of the ionization and recombination rates result in significant differences with respect to the earlier calculations, e.g. of \citet{Mazzotta98}.

The availability of accurate atomic data are of crucial importance in correct determination of the properties of the radiating astrophysical environment, and the solar corona in particular. In this paper, we present up-to-date calculations of the ionization and recombination rates (Sect. \ref{Sect:3}), and ionization equilibria (Sect. \ref{Sect:4}) for $\kappa$-distributions for elements from H to Zn. Such calculations are necessary for diagnostics of the $\kappa$-distributions in both the solar transition region and the corona, as well as subsequent calculations of the radiative losses \citep{Dudik11} or responses of various extreme-ultraviolet or X-ray filters \citep{Dudik09} used both to model and observe these portions of the solar atmosphere, with potential applications to other astrophysical environments.


%
%
\section{The Non-Maxwellian $\kappa$-distributions}
\label{Sect:2}

\begin{figure}
	\centering
	\includegraphics[width=8.6cm]{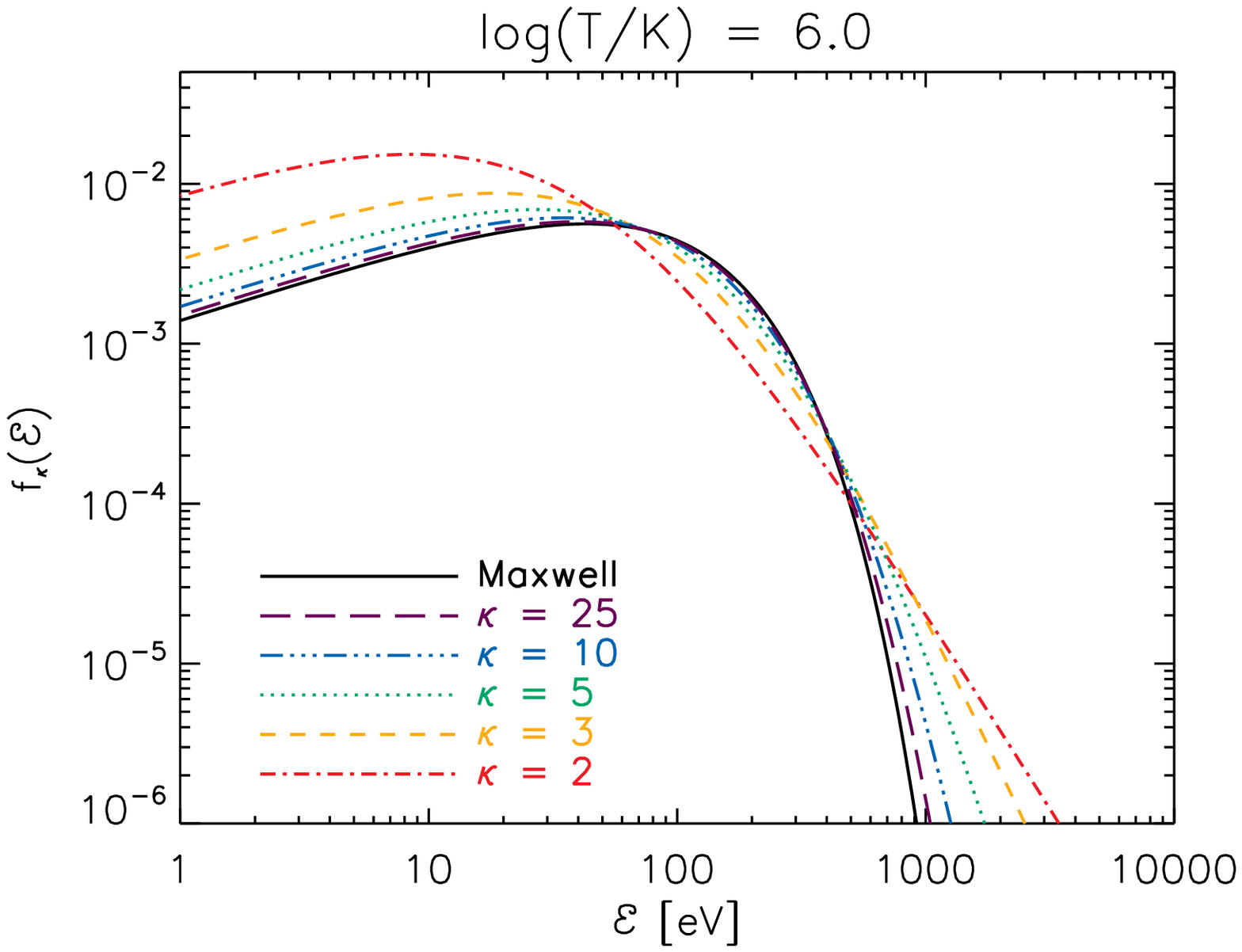}
	\includegraphics[width=8.6cm]{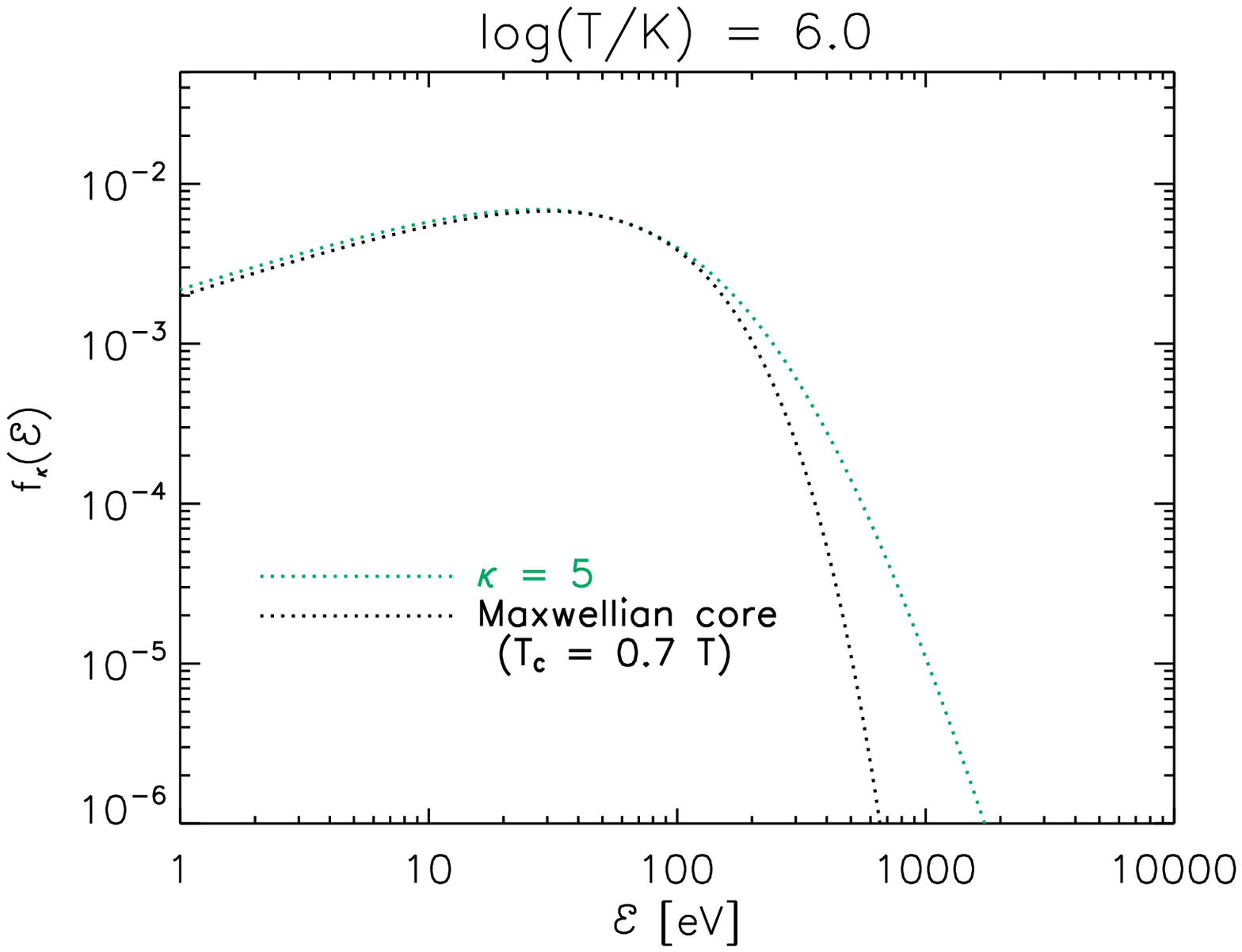}
\caption{\textit{Top}: The $\kappa$-distributions with $\kappa$\,=\,2, 3, 5, 10, 25 and the Maxwellian distribution plotted for log($T$/K)\,=\,6.0. \textit{Bottom}: The $\kappa$\,=\,5 distribution and the approximation of its core with Maxwellian distribution with lower $T_\mathrm{C}$. Colors and linestyles correspond to different values of $\kappa$. (A color version of this figure is available in the online journal.)
\label{Fig:kappa}}
\end{figure}

\notetoeditor{The Fig. \ref{Fig:kappa} should appear in this section, or on the same page.}

The $\kappa$-distributions of electron kinetic energies ${\cal E}$ represent a family of non-Maxwellian distributions characterized by two parameters, $\kappa$ and $T$
\begin{equation}
	f_{\kappa}({\cal E})d{\cal E} = {\cal A}_{\kappa}~ \frac{2}{\pi^{1/2} (k_\mathrm{B}T)^{3/2}}~ \frac{{\cal E}^{1/2}d { \cal E} } { (1 + \frac { {\cal E}}{( \kappa - 1.5) kT})^{ \kappa + 1 }}\,,
	\label{Eq:kappa}
\end{equation}
where the $ {\cal A}_{\kappa} = \Gamma ( \kappa + 1 )/\left[ \Gamma (\kappa -0.5) ( \kappa - 1.5 )^{3/2}\right]$ is the normalization constant, $k_\mathrm{B}$\,=\,1.38 $\times 10^{-23}$ J\,kg$^{-1}$ is the Boltzmann constant, and $\kappa$\,$\in$\,$\left(3/2,+\infty\right)$, $T$\,$\in$\,$\left(0,+\infty\right)$. We note that the definition of $\kappa$-distributions in Eq. (\ref{Eq:kappa} \textit{top}) corresponds to the $\kappa$-distributions of the second kind \citep[e.g.,][]{Livadiotis09}.

The shape of the distribution is controlled by the parameter $\kappa$. Maxwellian distribution is recovered for $\kappa$\,$\to$\,$+\infty$, while the departures from the Maxwellian increase with $\kappa$\,$\to$\,3/2. The departures from the Maxwellian distribution with decreasing $\kappa$ include increase of the number of particles in the high-energy tail as well as increase in the relative number of low-energy electrons (Fig. \ref{Fig:kappa} \textit{top}). However, the mean energy $\left<{\cal E}\right>$\,=\,3/2$k_\mathrm{B}T$ of the distribution does not depend on $\kappa$. This allows for calculation of all quantities depending on the mean energy of the distribution, e.g. pressure \citep{Dzifcakova06}. The parameter $T$ has in the frame of nonextensive statistics \citep{Tsallis88,Tsallis09} an analogous meaning as thermodynamic temperature in the Boltzmann-Gibbs statistics. The reader is referred e.g. to the work of \citet{Livadiotis09,Livadiotis10} for details.

While the shape of the $\kappa$-distribution differs from the Maxwellian with the same $T$ at all energies $E$, the core of the $\kappa$-distribution can be approximated by a Maxwellian distribution with lower $T_\mathrm{C}$\,=\,$T (\kappa -3/2)/\kappa$ \citep{Oka13}. An example for $\kappa$\,=\,5 and $T$\,=\,1\,MK is shown in Fig. \ref{Fig:kappa} \textit{bottom}, where the approximating Maxwellian distribution has $T_\mathrm{C} = 0.7T$ and has been scaled by the factor $C$
\begin{equation}
	C = 2.718 \frac{\Gamma(\kappa+1)}{\Gamma(\kappa -1/2)} \kappa^{-3/2} \left(1 +\frac{1}{\kappa}\right)^{-(\kappa+1)}\,
	\label{eq:scaling_kappa}
\end{equation}
\citep{Oka13}, which is $\approx$0.84 for $\kappa$\,=\,5. The difference between the $f_\kappa(T)$ and $f_\mathrm{Maxw}(T_\mathrm{C})$ are then mainly in the pronounced high-energy tail. This shows that the $\kappa$ distributions offer straightforward approximation of situations where a non-Maxwellian high-energy tail is present.

%
%
%
\section{Ionization and Recombination Rates}
\label{Sect:3}

To calculate the ionization and recombination rates for $\kappa$-distributions, we use the atomic data available through the CHIANTI database for astrophysical spectroscopy of optically thin plasmas \citep{Dere97,Landi13}. The analytical functional form of the $\kappa$-distributions allows for relatively simple direct integration of the ionization cross-sections (Sect \ref{Sect:3.2}). However, the recombination cross-sections are not contained in CHIANTI. These then have to be reverse-engineered from the Maxwellian recombination rates using assumptions detailed in Sect. \ref{Sect:3.3}.

We note that the CHIANTI database allows for the treatment of non-Maxwellian distributions only if these can be represented by a series of individual Maxwellian distributions (with different $T$s). The technique for calculation of ionization and recombination rates presented here can in principle be extended for any type of particle distribution, not only $\kappa$-distributions.

\subsection{Atomic Data}
\label{Sect:3.1}

The CHIANTI atomic database since version 6 \citep{Dere09} contains continually updated ionization equilibrium for the Maxwellian distribution. This ionization equilibrium utilizes the cross-sections for direct ionization and autoionization, and the corresponding rate coefficients from the work of \citet{Dere07}. Dielectronic and radiative recombination coefficients for the H to Al and Ar isoelectronic sequences are taken from the works of N. Badnell and colleagues, listed at \textit{http://amdpp.phys.strath.ac.uk/tamoc/DATA/} \citep{Badnell03,Colgan03,Colgan04,Mitnik04,Badnell06A,Altun05,Altun06,Altun07,Zatsarinny05A,Zatsarinny05B,Zatsarinny06,Bautista07,Nikolic10,Abdel-Naby12}. For the Si to Mn isoelectronic sequences, the recombination rates are based on the works of \citet{Shull82}, \citet{Nahar96,Nahar97}, \citet{Nahar01}, \citet{Mazzotta98} and \citet{Mazzitelli02}. The reader is referred to \citet{Dere09} for details. In the calculations presented in this paper, all atomic data for ionization and 
recombination correspond to the ones used to produce the \textit{chianti.ioneq} file in the CHIANTI v7.1. We note that ionization equilibria were published also by other authors \citep[e.g.][]{Bryans09}, but these use mostly the same atomic data as the CHIANTI database.



\subsection{Ionization Rates}
\label{Sect:3.2}

The ionization rates have been calculated by a numerical integration of the ionization cross section, $\sigma_{i}$, available in CHIANTI database
\begin{equation}
R_i=<\sigma_{i} v>\,=\int_{0}^{\infty}\sigma_{i} \left( \frac{2\cal E}{m}\right)^{1/2}f_{\kappa}({\cal E})d{\cal E},
\end{equation}
for $\kappa$\,=\,2,\,3,\,5,\,7,\,10, 25, and 33. Each numerical integral is calculated as a sum of individual integrals splitted according to the ionization and auto-ionizaton energies.

\begin{figure}
	\centering
	\includegraphics[width=8.0cm]{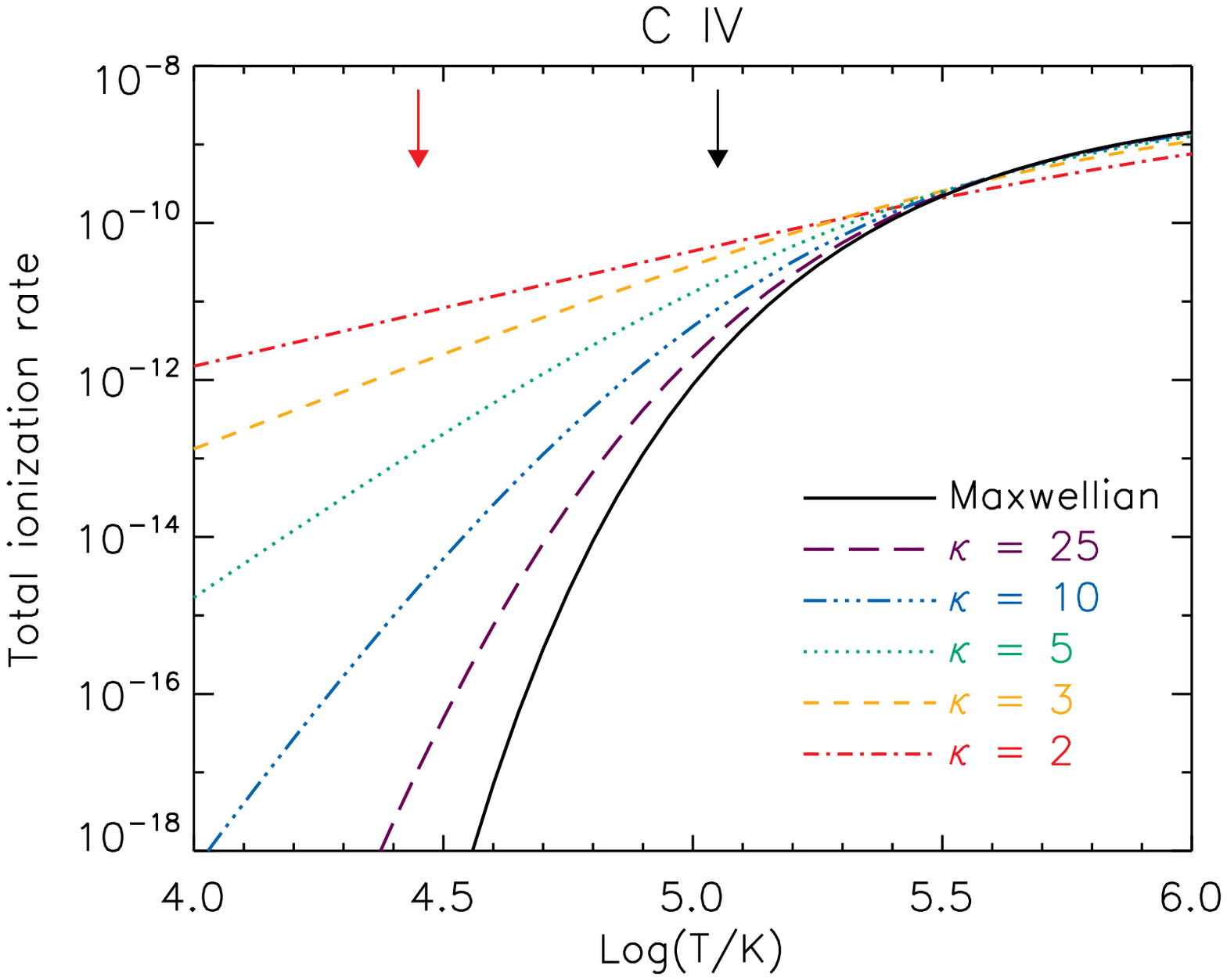}
	\includegraphics[width=8.0cm]{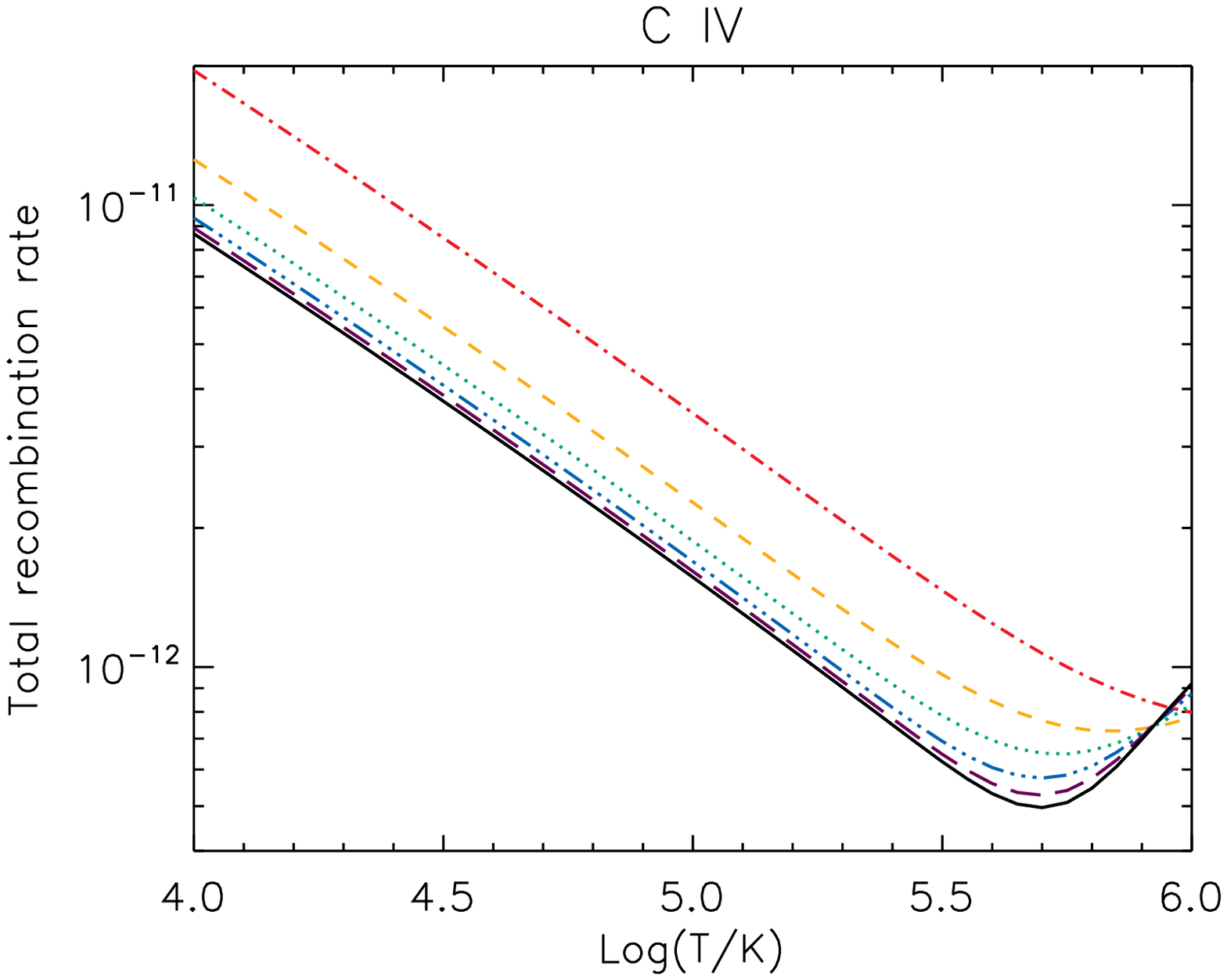}
	\includegraphics[width=8.0cm]{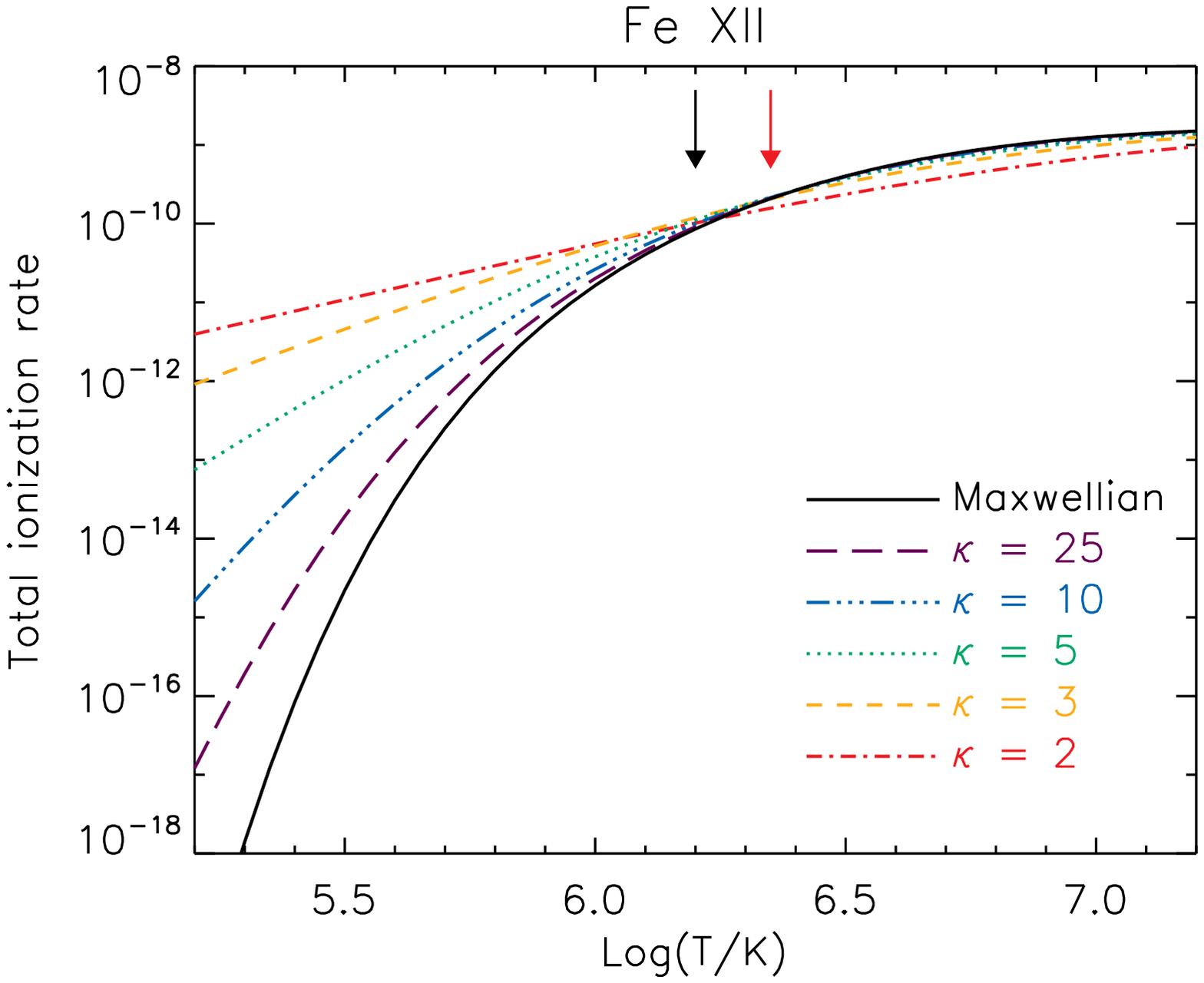}
	\includegraphics[width=8.0cm]{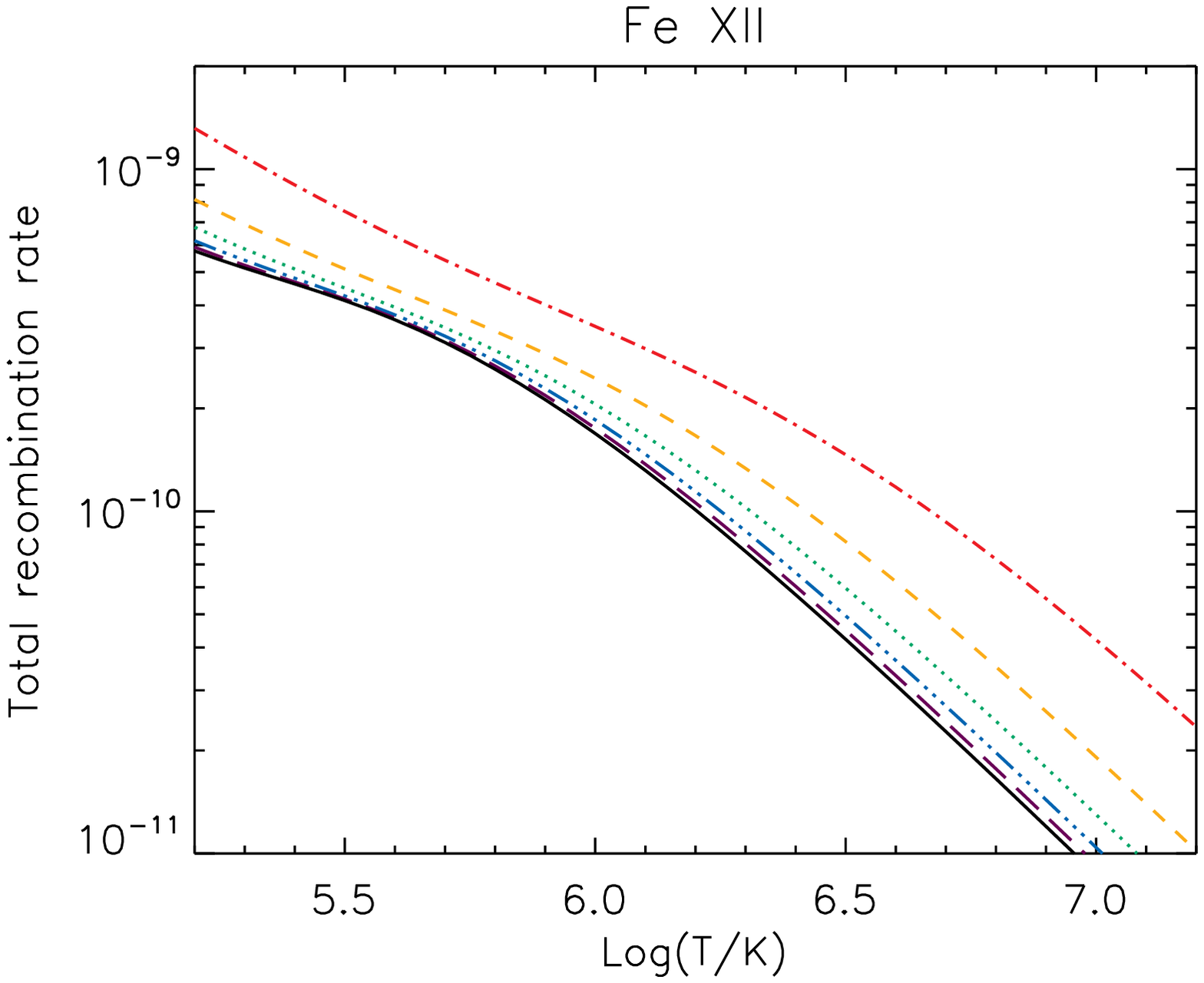}
	\includegraphics[width=8.0cm]{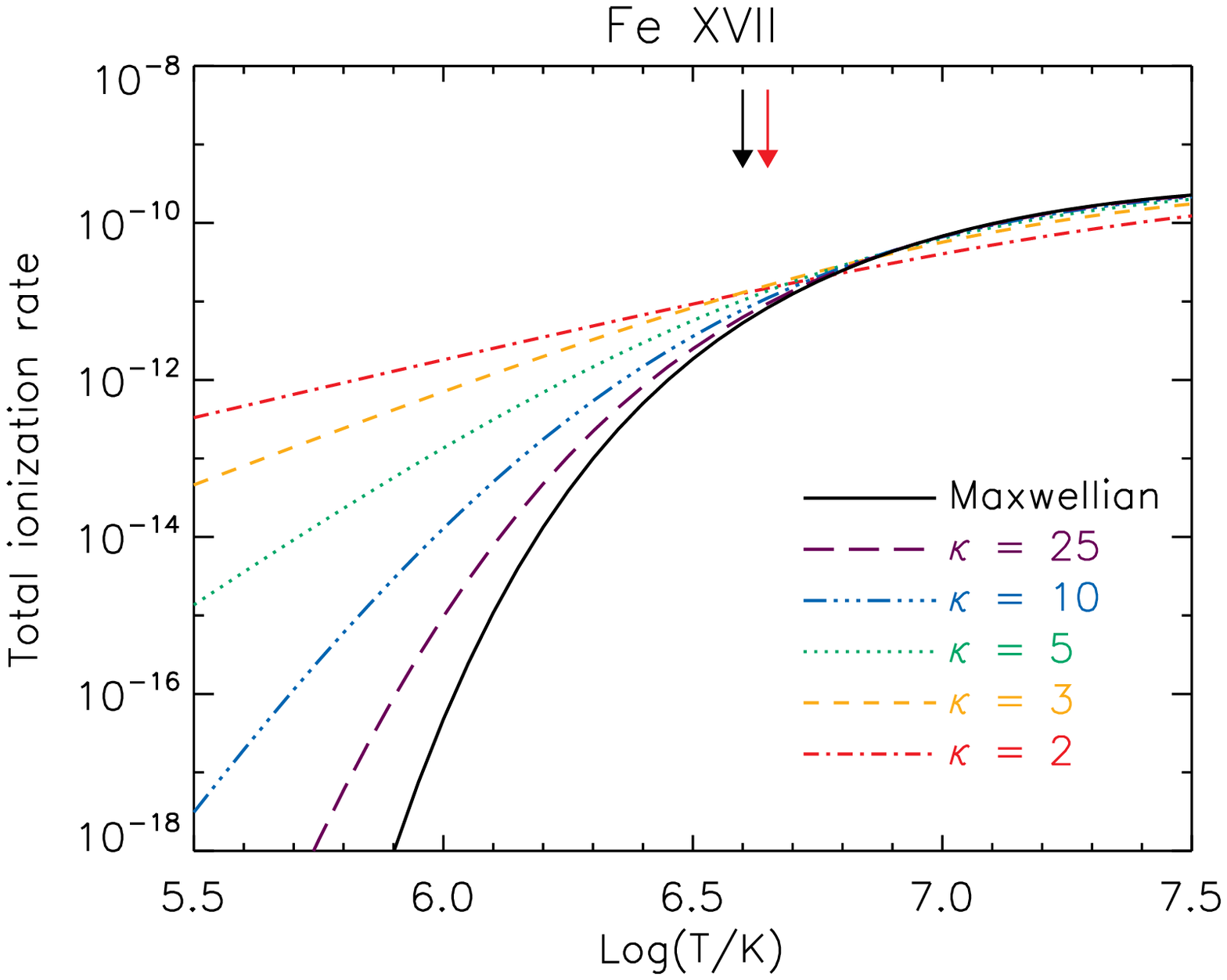}
	\includegraphics[width=8.0cm]{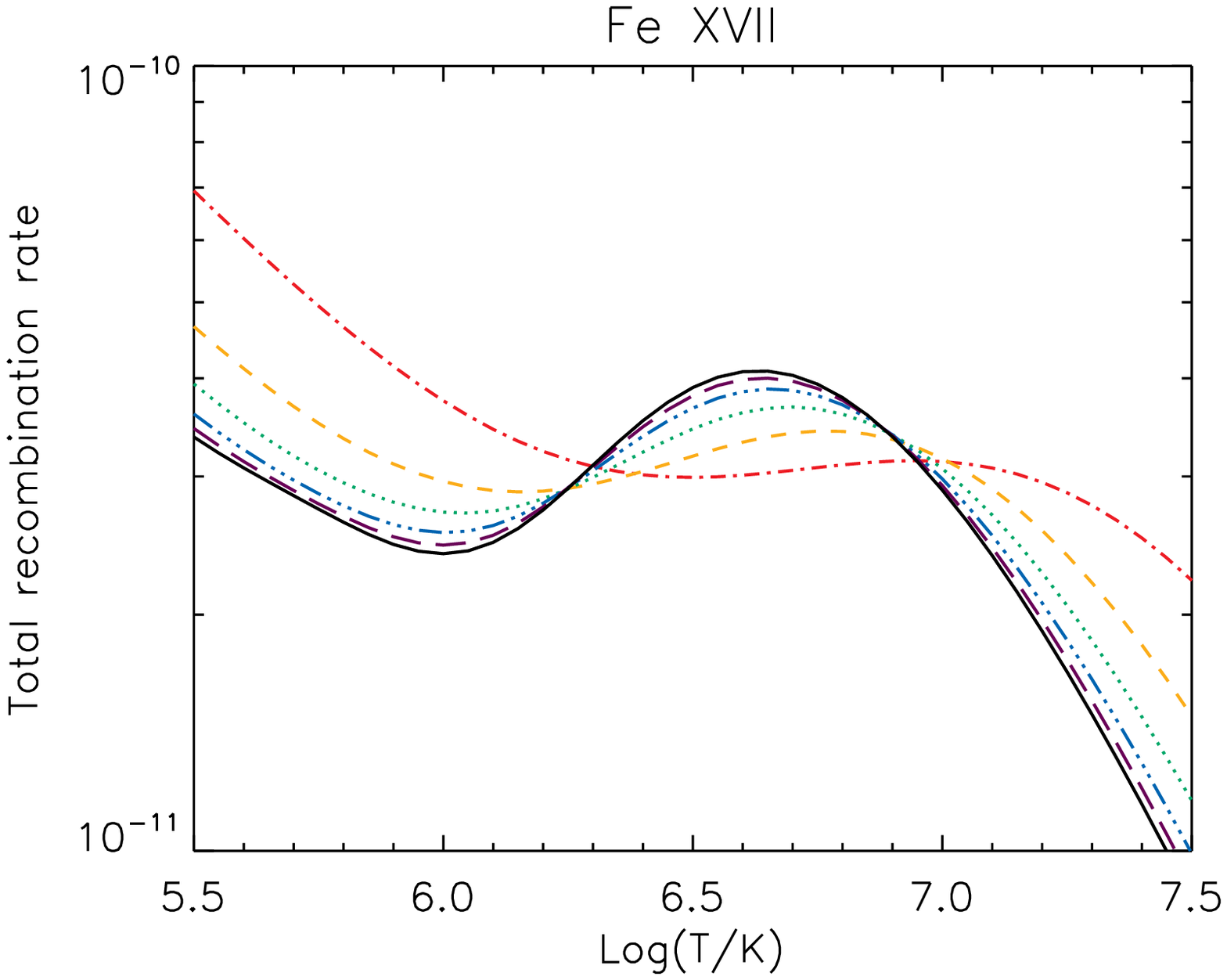}
\caption{Total ionization and recombination rates for \ion{C}{4} (\textit{top row}), \ion{Fe}{12} (\textit{middle row}) and \ion{Fe}{17} (\textit{bottom row}). Different linestyles correspond to different $\kappa$. Black and red arrows denote maximum of the relative ion abundance for the Maxwellian and $\kappa$\,=\,2 distributions, respectively. (A color version of this figure is available in the online journal.)
\label{Fig:rates}}
\end{figure}

Typical changes in behaviour of the direct ionization and autoionization rates with $\kappa$-distributions are shown in Fig. \ref{Fig:rates} for the ions \ion{C}{4}, \ion{Fe}{12}, and \ion{Fe}{17} formed at transition region, quiet coronal, and flare conditions, respectively. For lower $\kappa$, the temperature dependence of the ionization rates is much flatter, with lower maxima. Typically, for temperatures lower than those at which the ion abundance peaks (denoted by arrow in Fig. \ref{Fig:rates}, see also Sect. \ref{Sect:4}), the $\kappa$-distributions result in increase of the ionization rates by up to several orders of magnitude with respect to the Maxwellian distribution. These deviations from Maxwellian ionization rates increase with decreasing $\kappa$.

\subsection{Recombination Rates}
\label{Sect:3.3}

Since the individual recombination cross-sections are not available in CHIANTI, the calculation of the recombination rates for $\kappa$-distributions was performed using the method of \citet{Dzifcakova92}, used also by \citet{Wannawichian03}. This method allows for calculation of the recombination rates using approximations to the rates for the Maxwellian distribution.

The cross-section $\sigma_\mathrm{RR}$ for the radiative recombination is assumed to have a power-law dependence on energy \citep{Osterbrock74}
\begin{equation}
	\sigma_\mathrm{RR}({\cal E})=C_\mathrm{RR}/{\cal{E}}^{\eta+0.5}\,,
	\label{Eq:sigma_RR}
\end{equation}
where $C_\mathrm{RR}$ is a constant and $\eta+0.5$ is a power-law index. Subsequently, the radiative recombination rate for the $\kappa$-distribution is of the form:
\begin{equation}
 	R_\mathrm{RR}^{\kappa}=\frac{4C_\mathrm{RR}}{(2\pi m)^{1/2}}\frac{\Gamma(\kappa+\eta-0.5)(\kappa-1.5)^{\eta}}{\Gamma(\kappa-0.5)}\frac{\Gamma(1.5-\eta)}{(k_\mathrm{B}T)^{\eta}}\,,
	\label{Eq:R_RR_kp}
\end{equation}
while the radiative recombition rate for the Maxwellian distribution is
\begin{equation}
 	R_\mathrm{RR}^\mathrm{Maxw}=\frac{4C_\mathrm{RR}}{(2\pi m)^{1/2}}\frac{\Gamma(1.5-\eta)}{(k_\mathrm{B}T)^{\eta}}\,.
	\label{R_RR_Mxw}
\end{equation}
Therefore, it holds that
\begin{equation}
  	R_\mathrm{RR}^{\kappa}=R_\mathrm{RR}^\mathrm{Maxw}\frac{\Gamma(\kappa+\eta-0.5)(\kappa-1.5)^{\eta}}{\Gamma(\kappa-0.5)}\,.
	\label{Eq:R_RR}
\end{equation}
The level of error introduced by this approximation is typically several per cent, i.e., lower than the error of the atomic data themselves.

For the dielectronic recombination, following approximation has been taken \citep{Dzifcakova92}
\begin{equation}
	R_\mathrm{DR}^{\kappa}=\frac{A_{\kappa}}{T^{1/3}}\sum_i \frac{a_i}{(1+t_i/(\kappa-1.5)T)^{(\kappa+1)}}\,,
	\label{Eq:R_DR_kp}
\end{equation}
where parameters $a_i$ and $t_i$ are the same as in similar expressions for the Maxwellian distribution
\begin{equation}
	R_\mathrm{DR}^\mathrm{Maxw}=\frac{1}{T^{1/3}}\sum_i a_i \exp{\left(-t_i/T\right)} \,,
	\label{Eq:R_DR_Mxw}
\end{equation}
where the coefficients $a_i$ and $t_i$ are provided within the CHIANTI database. The precision of this approximation is given by the magnitude of the second-order terms in the expansion of the coefficients $a_i$ and $t_i$ into series \citep[Eqs. 37--44 in ][]{Dzifcakova92}.

The typical behavior of the total recombination rates ($R_\mathrm{RR} + R_\mathrm{DR}$) for $\kappa$-distributions and for Maxwellian distribution is shown in Fig. \ref{Fig:rates}. It can be seen that the radiative recombination rate increases with decrease of $\kappa$. This is a result of increasing number of low energy electrons for the $\kappa$-distributions (Fig. \ref{Fig:kappa}), which dominate the recombination processes. The local change of slope of total recombination rate is caused by the contribution of dielectronic recombination. For some ions, e.g. \ion{Fe}{17}, the contribution of dielectronic recombination is dominant in temperature interval where these ions have a non-negligible abundance.

%
%
\section{The Ionization Equilibrium}
\label{Sect:4}

\begin{figure}
	\centering
	\includegraphics[width=17.0cm]{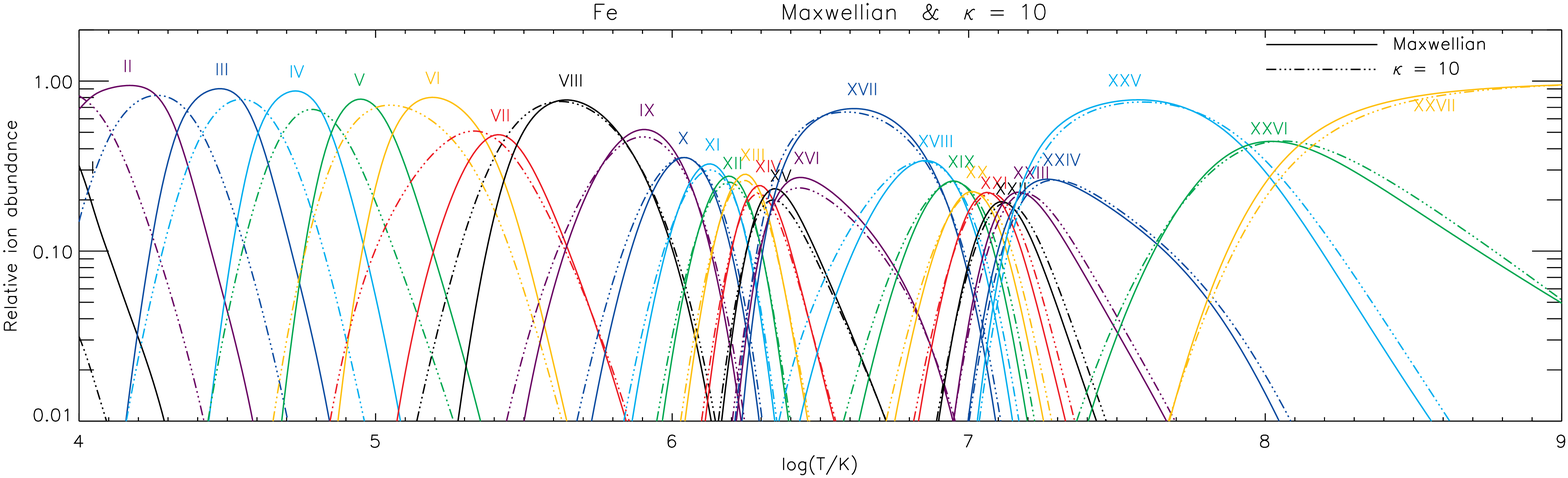}
	\includegraphics[width=17.0cm]{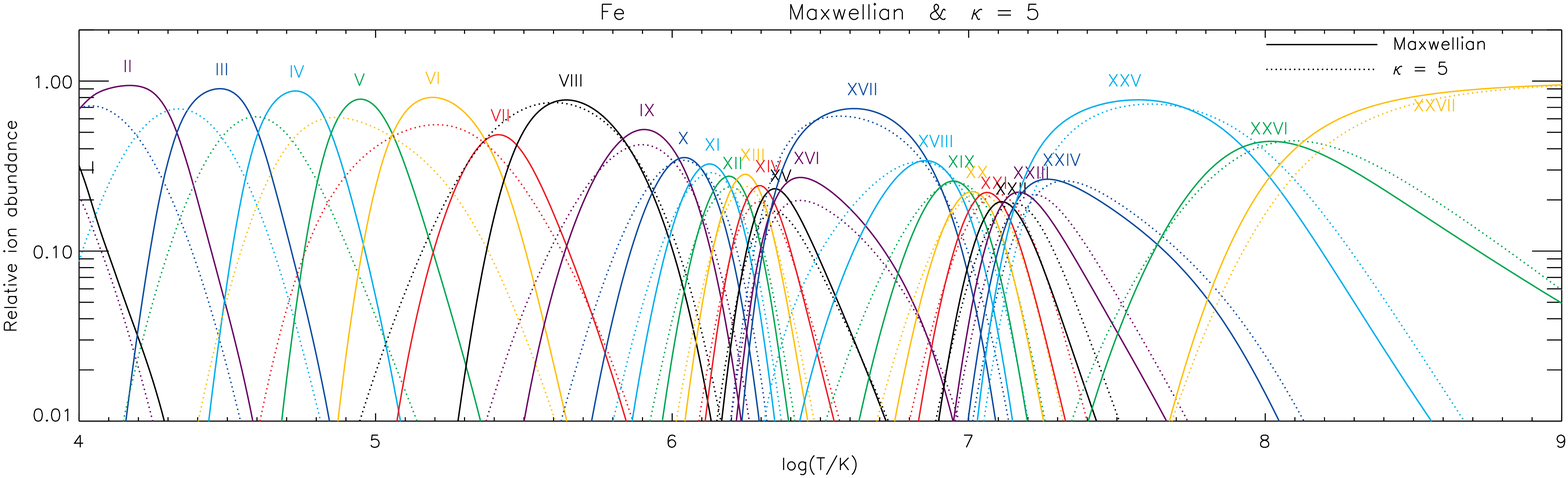}
	\includegraphics[width=17.0cm]{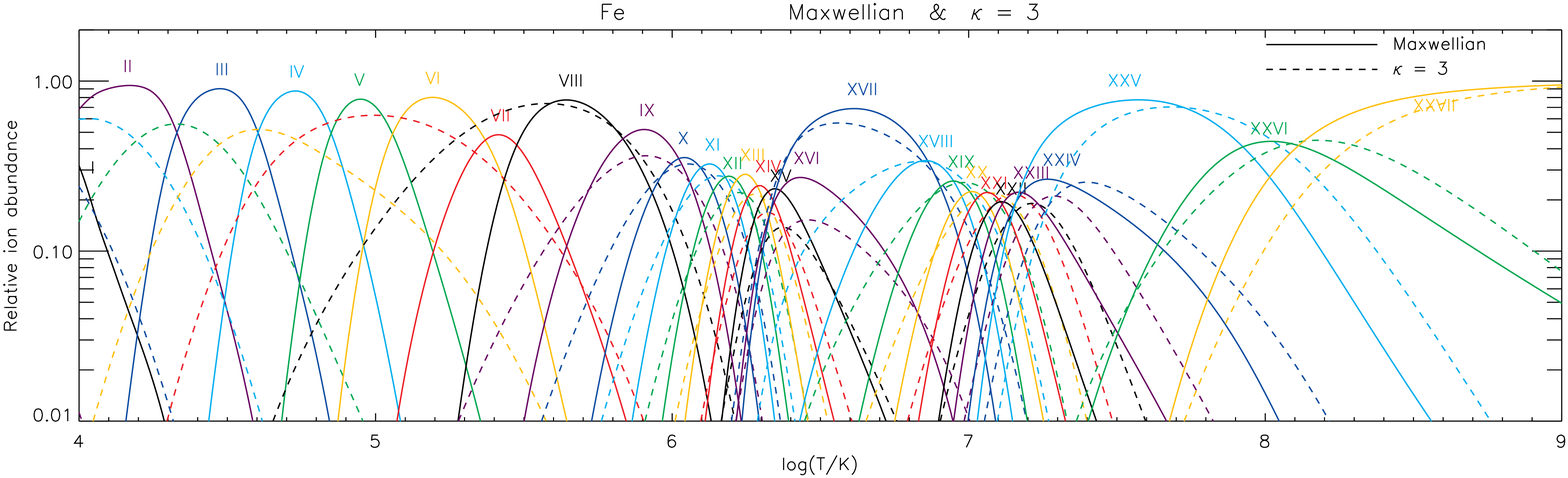}
	\includegraphics[width=17.0cm]{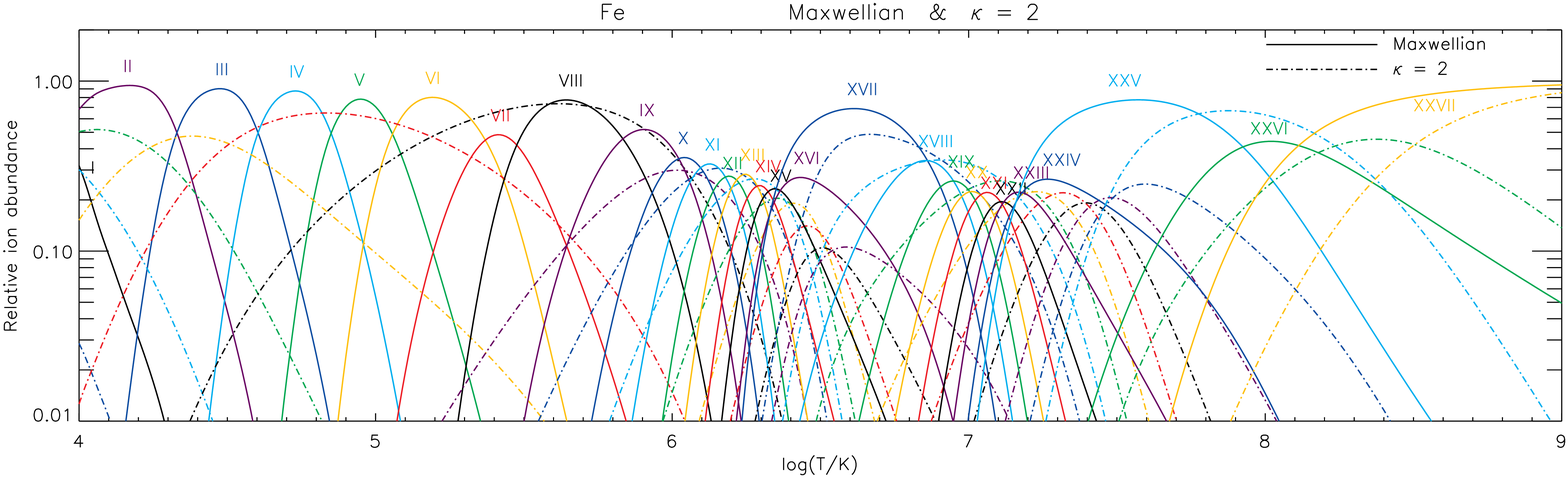}
\caption{Changes in the ionization equilibrium with $\kappa$ for iron. \textit{Top}: $\kappa$\,=\,10; \textit{Second row}: $\kappa$\,=\,5; \textit{Third row}: $\kappa$\,=\,5; \textit{Bottom row}: $\kappa$\,=\,2. Individual ionization stages are indicated. (A color version of this figure is available in the online journal.)
\label{Fig:ioneq_Fe}}
\end{figure}

\begin{figure}
	\centering
	\includegraphics[width=17.0cm]{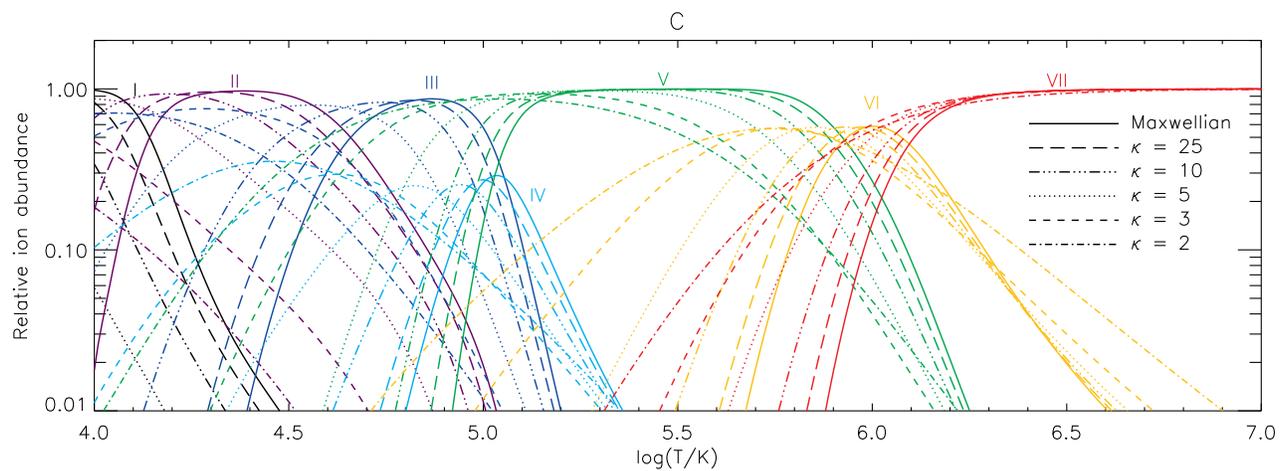}
\caption{Ionization equilibrium for Carbon. Different linestyles correspond to different $\kappa$. (A color version of this figure is available in the online journal.)
\label{Fig:ioneq_C}}
\end{figure}

\notetoeditor{The Figs. \ref{Fig:ioneq_Fe} and \ref{Fig:ioneq_C} should span the entire width of the page (about 18 cm) and should be located somewhere near the text of this section.}

Calculations of the collisional ionization equilibrium assume that there are no temporal variations in plasma temperature $T$. In the coronal conditions, the resulting relative ion populations are given by the equilibrium between the direct collisional ionization with auto-ionization and the radiative and dielectronic recombination. Three-body processes can be neglected at low electron densities typical in the solar corona \citep{Phillips08}. The radiative field is also usually assumed to be too weak, i.e., photoionization can be neglected as well. However, we note that photoionization may be important for some transition-region ions with low ionization thresholds, but the effect will vary depending on the distance from the radiation field (e.g., the solar photosphere). 

The ionization equilibrium for $\kappa$-distributions with $\kappa$\,=\,2,\,3,\,5,\,7,\,10, 25, and 33 was calculated for ions of astrophysical interests, ranging from H ($Z$\,=\,1) to Zn ($Z$\,=\,30). Previous calculations of \citet{Dzifcakova92}, \citet{Dzifcakova02}, and \citet{Wannawichian03} involved only 12 most abundant elements. The current calculations are performed for the same set of temperatures as the calculations in the CHIANTI database in the \textit{chianti.ioneq} file. I.e., the temperature spans the interval of log($T$/K)\,$\in$\,$\left<4,9\right>$ with the step of $\Delta$log($T$/K)\,=\,0.05. The calculations are available in the same format as the \textit{chianti.ioneq} file and are easily readable using the routines \textit{read\_ioneq.pro} and \textit{plot\_ioneq.pro} available in CHIANTI running under SolarSoftware in IDL. More information on the CHIANTI \textit{.ioneq} file format is provided in Appendix \ref{Appendix:A}.

Examples of the current ionization equilibrium calculations for iron are in Fig. \ref{Fig:ioneq_Fe} and for carbon in Fig. \ref{Fig:ioneq_C}. The typical behavior is that the lower $\kappa$, the flatter the ionization peaks. In addition, ionization peaks can be shifted to lower or higher $T$, depending on $\kappa$, $T$ and the individual ion. This means that a given relative ion abundance can be formed at a wider range of $T$ for a $\kappa$-distribution, with different peak formation temperature. Such behavior was already reported by \citet{Dzifcakova92}, \citet{Dzifcakova02} and \citet{Wannawichian03}. Typically, the shifts of Fe and C ion abundance peaks are to lower $T$ with respect to the Maxwellian ionization equilibrium if log($T$/K)\,$\lessapprox$\,5.5. At coronal temperatures, the \ion{Fe}{9} -- \ion{Fe}{16} ions are shifted to higher $T$. An interesting example is the ion \ion{Fe}{17}, whose ionization peak is shifted to lower $T$ for $\kappa$\,$\
lessapprox$\,3, but to higher $T$ for $\kappa$\,=\,2 (Fig. \ref{Fig:ioneq_Fe} \textit{bottom}).

\begin{figure}
	\centering
	\includegraphics[width=17.0cm]{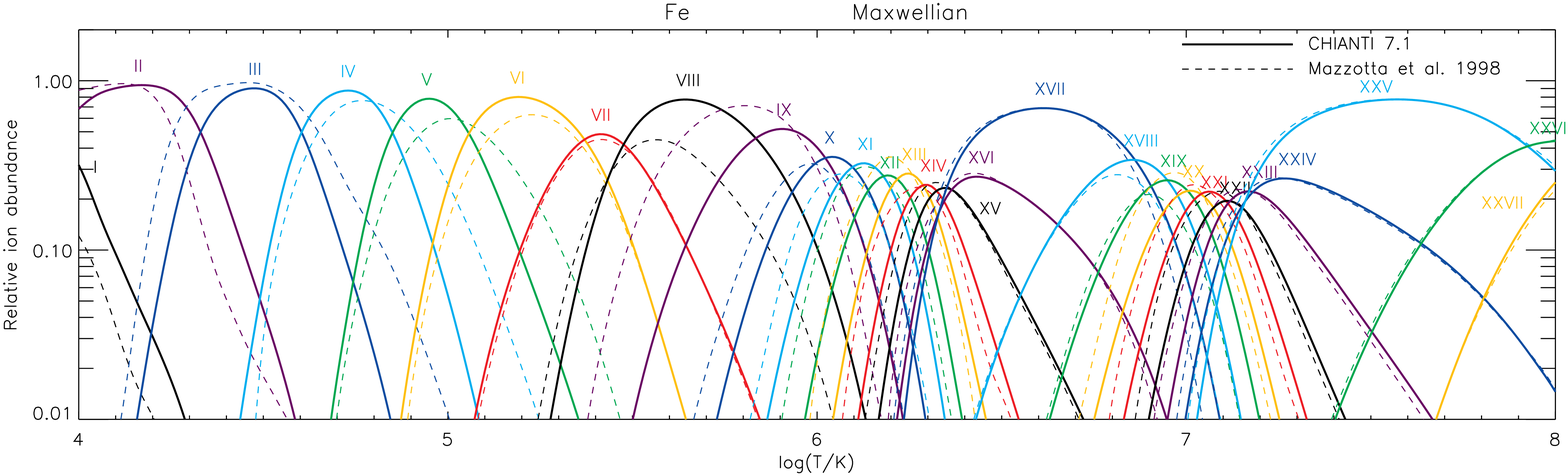}
	\includegraphics[width=17.0cm]{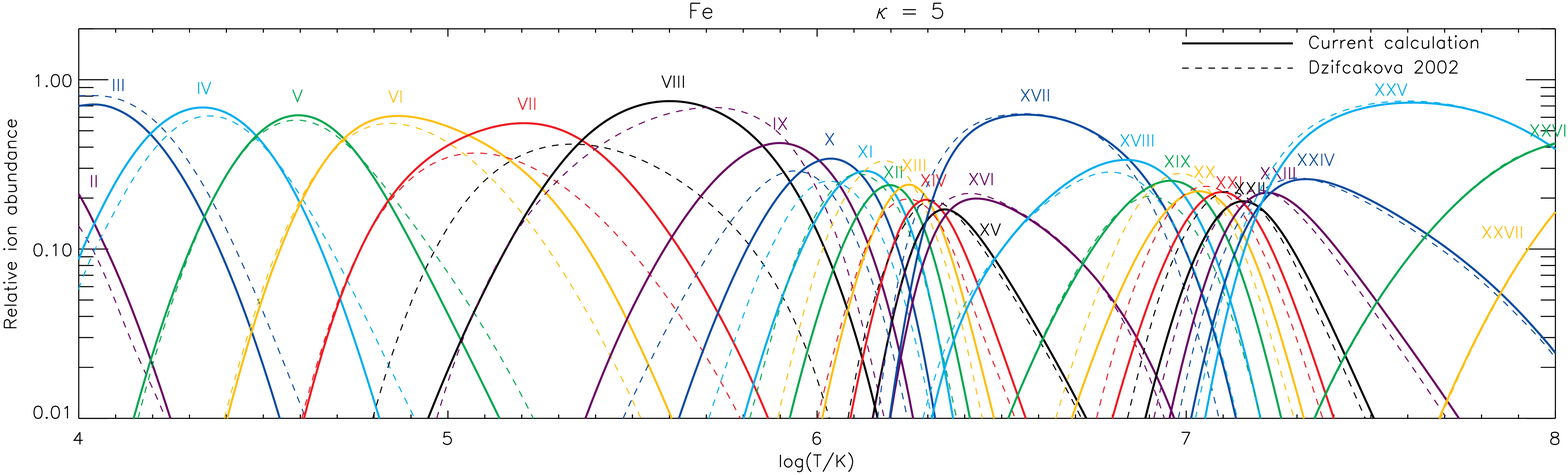}
	\includegraphics[width=17.0cm]{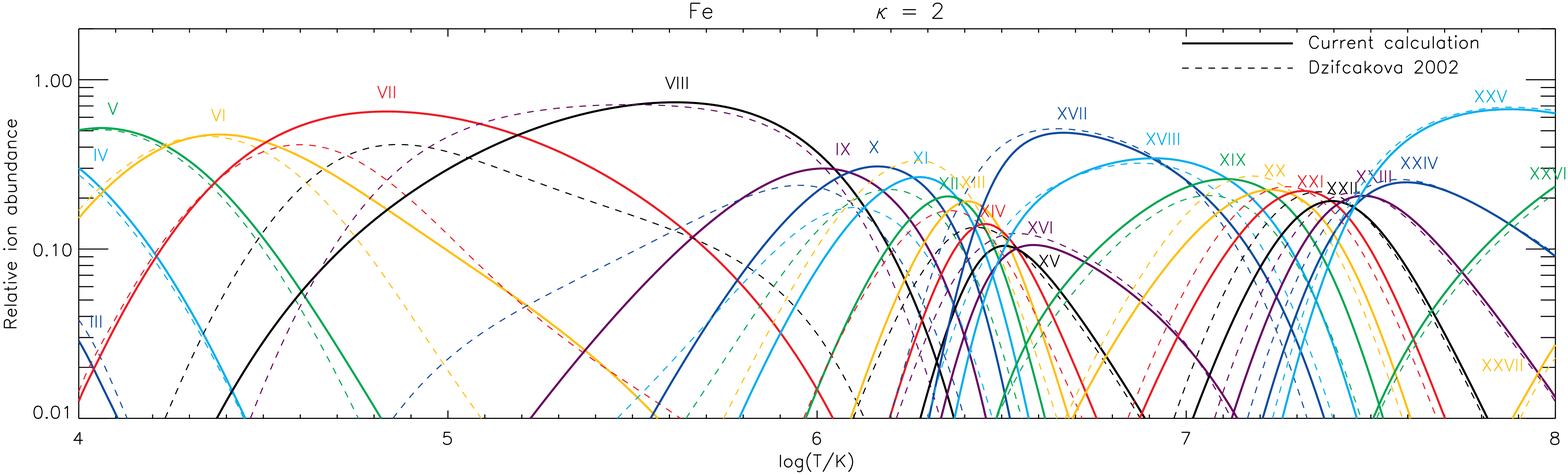}
\caption{Comparison of the current, updated calculations with the previous ones. \textit{Top}: Comparison between the ionization equilibrium according to \citet{Dere09} and \citet{Mazzotta98}. \textit{Middle} and \textit{Bottom}: Comparison of the current calculations with the ones of \citet{Dzifcakova02} for $\kappa$=\,5 and\,2, respectively. Individual ionization stages are indicated. (A color version of this figure is available in the online journal.)
\label{Fig:ioneq_Fe_compare}}
\end{figure}

There are a number of differences with respect to the earlier calculations of \citet{Dzifcakova02}, which used the atomic data corresponding to those of \citet{Mazzotta98}. These differences are illustrated in Fig. \ref{Fig:ioneq_Fe_compare}. In general, the lower the value of $\kappa$, the greater the differences with respect to the previous calculations. The most conspicuous examples are the \ion{Fe}{9}, which is shifted to slightly higher $T$ instead to lower $T$ as in the previous calculations of \citet{Dzifcakova02}. The \ion{Fe}{12} and \ion{Fe}{13} ions are shifted to higher $T$ (up to 0.1--0.15 dex). We note that the changes in the ionization equilibria due to the updated atomic data will reflect e.g. on changes in the total radiative losses \citep{Dudik11} and will also modify the proposed diagnostic methods for $\kappa$-distributions \citep{Dzifcakova10,Mackovjak13}.

\begin{figure}
	\centering
	\includegraphics[width=8.0cm]{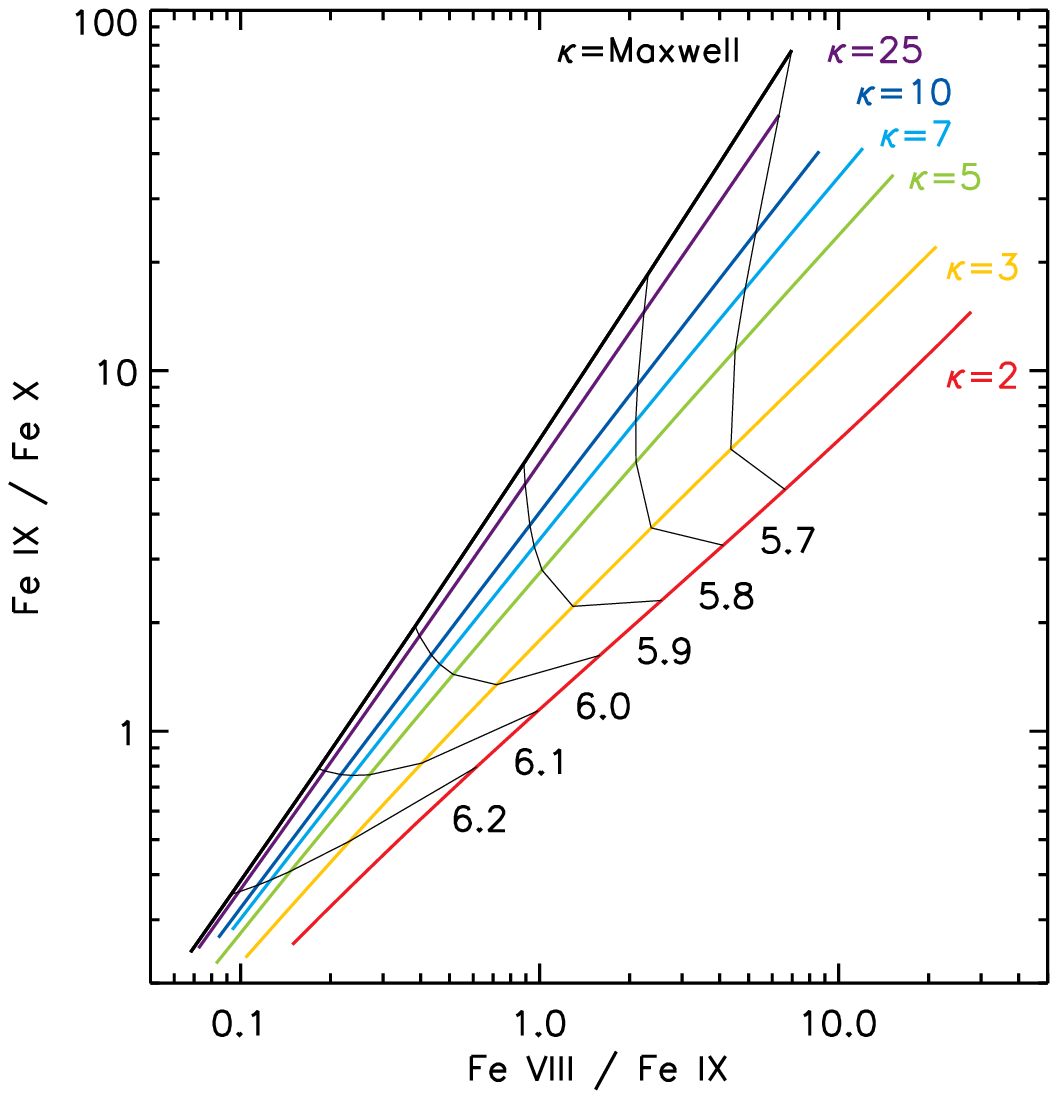}
	\includegraphics[width=8.0cm]{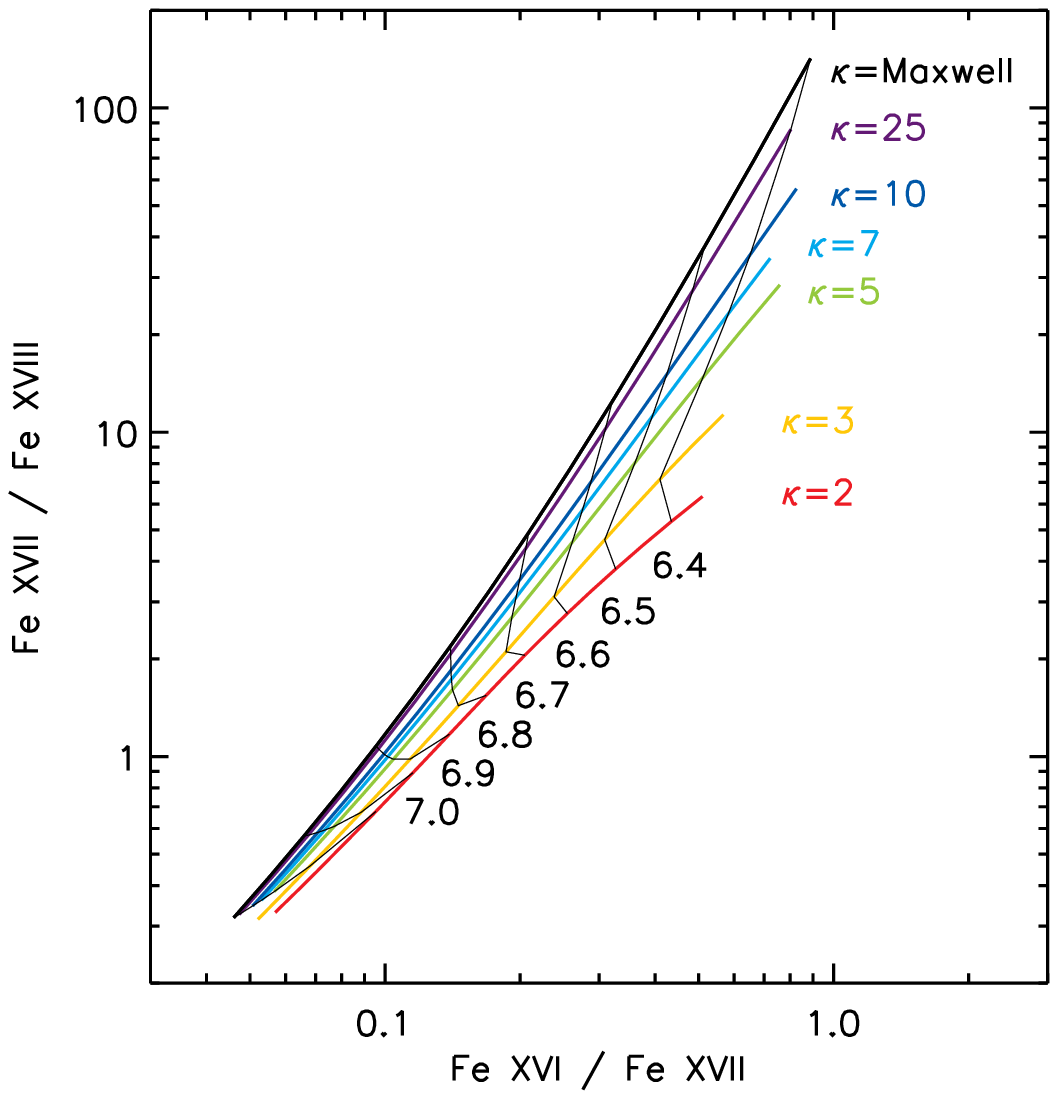}
\caption{Diagnostics of $\kappa$ from ratios of ion abundances. The values of $\kappa$ for each line are indicated. Thin lines represent isotherms for the given value of log($T$/K). (A color version of this figure is available in the online journal.)
\label{Fig:diag}}
\end{figure}

Ratios of relative abundances of individual ions can be used to diagnose the value of $\kappa$ and simultaneously $T$. Such diagnostics can be applied e.g. in the solar wind \citep[][]{Owocki83}. An example of diagnostic diagrams is given in Fig. \ref{Fig:diag}. We note that such diagrams cannot be directly applied on the observed line intensities unless the additional effect of $\kappa$-distributions on the excitation and deexcitation rates is considered. However, these diagrams provide quantification of the expected changes to ratios of line intensities due to ionization equilibrium. Ratios of lines involving different ionization stages provide better options for determining $\kappa$, as noted already by \citep{Dzifcakova10,Mackovjak13}.

%
%
\section{Summary}
\label{Sect:5}

Collisional ionization equilibrium calculations for $\kappa$-distributions and optically thin plasmas were performed for all ions of elements H to Zn. To do that, the latest available atomic data for ionization and recombination were used. The calculations are available in the form of ionization equilibrium files compatible with the CHIANTI database, v7.1.

For $\kappa$-distributions, the ionization peaks are in general flatter and can be shifted to lower or higher $T$. This means that for $\kappa$-distributions, individual ions are typically formed in a wider range of temperatures, with different peak formation temperatures. Typically, ions formed at transition region temperatures are shifted to lower $T$, while the majority of coronal iron ions are shifted to higher $T$. Comparison to previous calculations is provided. Due to updated atomic data calculations, several ions in the present calculations are shifted to different temperatures with respect to previous calculations of \citet{Dzifcakova02}. The effect of $\kappa$ on the ratios of ion abundances is documented.

The present calculations provide an accurate, necessary and useful tool for detection of astrophysical plasmas out of thermal equilibrium, where the distribution of particles is characterized by an enhanced power-law tail. The supplied files compatible with the CHIANTI database and software should greatly facilitate synthetization of both line and continuum spectra for optically thin astrophysical plasmas.

\acknowledgments
The authors are grateful to Dr. G. Del Zanna and Dr. H. E. Mason for helpful discussions. This work was supported by Grant Agency of the Czech Republic, Grant No. P209/12/1652, the project RVO:67985815, Scientific Grant Agency, VEGA, Slovakia, Grant No. 1/0240/11, and the bilateral project APVV CZ-SK-0153-11 (7AMB12SK154) involving the Slovak Research and Development Agency and the Ministry of Education of the Czech Republic. JD acknowledges support from the Comenius University Grant No. UK/11/2012. CHIANTI is a collaborative project involving the NRL (USA), RAL (UK), MSSL (UK), the Universities of Florence (Italy) and Cambridge (UK), and George Mason University (USA). CHIANTI is great spectroscopic database and software, and the authors are very grateful for its existence and availability.

\appendix
\section{CHIANTI \textit{.ioneq} file format}
\label{Appendix:A}

The \textit{.ioneq} files, used by the CHIANTI database and software to store data on the ionization equilibrium, are in essence formatted ASCII files. An example of their format is given in Table \ref{Table:1}. The file begins with two numbers, $N$, giving the number of temperature points $T_i$, $1 \le i \le N$, and $Z_\mathrm{max}$, denoting the maximum proton number $Z$ for which the ionization equilibrium is provided. The next line gives the tabulated temperatures log$(T_i/\mathrm{K})$. All other following lines begin with the $Z$ and a positive integer $+k$ denoting the roman numeral for the corresponding ion; e.g., 26 and 9 stands for \ion{Fe}{9}. The line then contains relative ion abundances $A_Z^{+k-1}(T_i)$ in floating-point precision for the tabulated temperatures.

Portion of the actual content of the \textit{kappa\_05.ioneq} file is given in Table \ref{Table:2}. There, the relative abundances of the ions \ion{Fe}{9}--\ion{Fe}{11} are listed for $\kappa$\,=\,5 and for log$(T/\mathrm{K})$\,=\,5.80 -- 6.15.

\notetoeditor{Font size has been decreased in Tables 1 and 2. The LaTeX commands /tiny  (Table 1) and /scriptsize (Table 2) should be removed in the print version of the manuscript.}

\clearpage
\begin{table}[!h]
\begin{center}
\tiny
\caption{ASCII format of the CHIANTI \textit{.ioneq} file. Explanation of the symbols is given in the text. 
\label{Table:1}}
\begin{tabular}{cccccccccccc}
\tableline\tableline
$N$	& $Z_\mathrm{max}$ & 					\\
\tableline
 	& 	& log$(T_1/\mathrm{K})$	& log$(T_2/\mathrm{K})$	& log$(T_3/\mathrm{K})$	& log$(T_4/\mathrm{K})$	& log$(T_5/\mathrm{K})$	& log$(T_6/\mathrm{K})$ & ... & log$(T_i/\mathrm{K})$ & ... & log$(T_N/\mathrm{K})$ \\		
\tableline
1	& 1	& $A_\mathrm{H}^{+0}(T_1)$ & $A_\mathrm{H}^{+0}(T_2)$ & $A_\mathrm{H}^{+0}(T_3)$ & $A_\mathrm{H}^{+0}(T_4)$ & $A_\mathrm{H}^{+0}(T_5)$ & $A_\mathrm{H}^{+0}(T_6)$ & ... & $A_\mathrm{H}^{+0}(T_i)$ & ... & $A_\mathrm{H}^{+0}(T_N)$ \\
1	& 2	& $A_\mathrm{H}^{+1}(T_1)$ & $A_\mathrm{H}^{+1}(T_2)$ & $A_\mathrm{H}^{+1}(T_3)$ & $A_\mathrm{H}^{+1}(T_4)$ & $A_\mathrm{H}^{+1}(T_5)$ & $A_\mathrm{H}^{+1}(T_6)$ & ... & $A_\mathrm{H}^{+1}(T_i)$ & ... & $A_\mathrm{H}^{+1}(T_N)$ \\
\tableline
2	& 1	& $A_\mathrm{He}^{+0}(T_1)$ & $A_\mathrm{He}^{+0}(T_2)$ & $A_\mathrm{He}^{+0}(T_3)$ & $A_\mathrm{He}^{+0}(T_4)$ & $A_\mathrm{He}^{+0}(T_5)$ & $A_\mathrm{He}^{+0}(T_6)$ & ... & $A_\mathrm{He}^{+0}(T_i)$ & ... & $A_\mathrm{He}^{+0}(T_N)$ \\
2	& 2	& $A_\mathrm{He}^{+1}(T_1)$ & $A_\mathrm{He}^{+1}(T_2)$ & $A_\mathrm{He}^{+1}(T_3)$ & $A_\mathrm{He}^{+1}(T_4)$ & $A_\mathrm{He}^{+1}(T_5)$ & $A_\mathrm{He}^{+1}(T_6)$ & ... & $A_\mathrm{He}^{+1}(T_i)$ & ... & $A_\mathrm{He}^{+1}(T_N)$ \\
2	& 3	& $A_\mathrm{He}^{+2}(T_1)$ & $A_\mathrm{He}^{+2}(T_2)$ & $A_\mathrm{He}^{+2}(T_3)$ & $A_\mathrm{He}^{+2}(T_4)$ & $A_\mathrm{He}^{+2}(T_5)$ & $A_\mathrm{He}^{+2}(T_6)$ & ... & $A_\mathrm{He}^{+2}(T_i)$ & ... & $A_\mathrm{He}^{+2}(T_N)$ \\
\tableline
\\
...	& ... 	& ... 	& ...	& ...	& ...	& ...	& ... & ... & ... & ... & ... \\
\\
\tableline
$Z$	& 1	& $A_Z^{+0}(T_1)$ & $A_Z^{+0}(T_2)$ & $A_Z^{+0}(T_3)$ & $A_Z^{+0}(T_4)$ & $A_Z^{+0}(T_5)$ & $A_Z^{+0}(T_6)$ & ... & $A_Z^{+0}(T_i)$ & ... & $A_Z^{+0}(T_N)$ \\
$Z$	& ... 	& ... 	& ...	& ...	& ...	& ...	& ... & ... & ... & ... & ... \\	
$Z$	& $+(k+1)$	& $A_Z^{+k}(T_1)$ & $A_Z^{+k}(T_2)$ & $A_Z^{+k}(T_3)$ & $A_Z^{+k}(T_4)$ & $A_Z^{+k}(T_5)$ & $A_Z^{+k}(T_6)$ & ... & $A_Z^{+k}(T_i)$ & ... & $A_Z^{+k}(T_N)$ \\
$Z$	& ... 	& ... 	& ...	& ...	& ...	& ...	& ... & ... & ... & ... & ... \\	
$Z$	& $+(Z+1)$	& $A_Z^{+Z}(T_1)$ & $A_Z^{+Z}(T_2)$ & $A_Z^{+Z}(T_3)$ & $A_Z^{+Z}(T_4)$ & $A_Z^{+Z}(T_5)$ & $A_Z^{+Z}(T_6)$ & ... & $A_Z^{+Z}(T_i)$ & ... & $A_Z^{+Z}(T_N)$ \\
\tableline
\\
...	& ... 	& ... 	& ...	& ...	& ...	& ...	& ... & ... & ... & ... & ... \\
\\	
\tableline
$Z_\mathrm{max}$ & ...	& ... 	& ...	& ...	& ...	& ...	& ... & ... & ... & ... & ... \\
$Z_\mathrm{max}$ & $+(Z_\mathrm{max}+1)$	& ... 	& ...	& ...	& ...	& ...	& ... & ... & ... & ... & ... \\ \tableline\tableline
\end{tabular}
\end{center}
\end{table}

\clearpage
 
\begin{table}[!h]
\begin{center}
\scriptsize
\caption{Example of the $\textit{kappa\_05.ioneq}$ file produced for $\kappa\,=\,5$. The table shows relative abundances of the ions \ion{Fe}{9} -- \ion{Fe}{11}.
\label{Table:2}}
\tabletypesize{\scriptsize}
\begin{tabular}{ccccccccccccc}
\tableline\tableline
101	& 30 & 					\\
\tableline
 	& 	& ... & 5.80 & 5.85 & 5.90 & 5.95 & 6.00 & 6.05 & 6.10 & 6.15 & ... \\		
\tableline
...	& ... 	& ... 	& ...	& ...	& ...	& ...	& ... & ... & ... & ... & ... \\	
\tableline
26	& ...	& ... 	& ...	& ...	& ...	& ...	& ... & ... & ... & ... & ... \\
26	& 9 & ...   & 0.357000   & 0.405000  & 0.422800  & 0.401500  & 0.342300  & 0.257800  & 0.167600  & 0.091260  & ... \\
26	& 10 & ...  & 0.0905900  & 0.144700  & 0.211400  & 0.279200  & 0.329100  & 0.340500  & 0.302200  & 0.223100  & ... \\
26	& 11 &	... & 0.00983700 & 0.0228500 & 0.0481900 & 0.0912400 & 0.153000  & 0.223400  & 0.277700  & 0.284900  & ... \\
26	& ...	& ... 	& ...	& ...	& ...	& ...	& ... & ... & ... & ... & ... \\
\tableline
...	& ... 	& ... 	& ...	& ...	& ...	& ...	& ... & ... & ... & ... & ... \\	
\tableline\tableline
\end{tabular}
\end{center}
\end{table}


%
%
\bibliographystyle{apj}                       
\bibliography{Ioneq}

\clearpage

%
\clearpage

%
\clearpage

%
\clearpage

%
\clearpage

%
\clearpage

%
\clearpage

\clearpage
\end{document}